\documentclass[aps,prd,email,preprint,showpacs,showkeys,preprintnumbers,amsmath,amssymb,nofootinbib,superscriptaddress]{revtex4}

\usepackage{color}
\usepackage{latexsym}
\usepackage{amsmath}
\usepackage{amssymb}
\usepackage{eufrak}
\usepackage{euscript}
\usepackage{pstricks}
\usepackage{graphics}
\usepackage{graphicx}
\usepackage{picture}
\def\[{\left\lbrack}
\def\]{\right\rbrack}

\def\({\left(}
\def\){\right)}

\newcommand{\bbe}{\begin{equation}}
\newcommand{\eee}{\end{equation}}
\newcommand{\eaa}{\end{eqnarray}}
\newcommand{\baa}{\begin{eqnarray}}

\def\ni{\noindent}

\def\no{\nonumber}

\def\uma{\rm 1\!\!\hskip 1 pt l}
\begin{document}

\pagestyle{myheadings}
\markright{New remarks on DFR noncommutative phase-space}

\title{\Large{New remarks on DFR noncommutative phase-space}}

\author{Everton M. C. Abreu} \email{evertonabreu@ufrrj.br}
\affiliation{Grupo de F\' isica Te\'orica e Matem\'atica F\' isica, Departamento de F\'{\i}sica,
Universidade Federal Rural do Rio de Janeiro,
BR 465-07, 23890-971, Serop\'edica, RJ, Brazil}
\affiliation{Departamento de F\'{\i}sica, ICE, Universidade Federal de Juiz de Fora,
36036-330, Juiz de Fora, MG, Brazil}
\author{M. J. Neves$^{a}$}\email{mariojr@ufrrj.br}
\affiliation{Grupo de F\' isica Te\'orica e Matem\'atica F\' isica, Departamento de F\'{\i}sica,
Universidade Federal Rural do Rio de Janeiro,
BR 465-07, 23890-971, Serop\'edica, RJ, Brazil}
\author{Vahid Nikoofard}\email{vahid@fisica.ufjf.br}
\affiliation{Departamento de F\'{\i}sica, ICE, Universidade Federal de Juiz de Fora,
36036-330, Juiz de Fora, MG, Brazil}

\date{\today}



\begin{abstract}
\noindent  The so-called canonical noncommutativity is based on a constant noncommutative parameter ($\theta$).   However, this formalism breaks Lorentz invariance and one way to recover it is to define the NC parameter as a variable, an extra coordinate of the system.  One approach that uses the variable $\theta$ was developed by Doplicher, Fredenhagen and Roberts (DFR) and hence, their phase-space is formed by $(x,p,\theta)$ with extra-dimensions.   In this work we have demonstrated precisely that this phase-space is incomplete because the variable $\theta$ requires an associated momentum and the so-called DFR phase-space is in fact formed by $(x, p, \theta, \pi)$, where $\pi$ is an useful object. One of the models used here to demonstrate this fact brought other interesting results.   We have used this complete phase-space to explain some undefined results in the $\theta$-variable literature.   Finally, we have shown the importance of this DFR-momentum since with it we could fill the gap that exist in $\theta$-variable results.   In other words, we have computed the field commutation relations of a QFT in this DFR phase-space.  The results obtained here match exactly with the postulated (not demonstrated) values that dwell in the DFR literature.
\end{abstract}


\maketitle

\pagestyle{myheadings}
\markright{New remarks on DFR noncommutative phase-space}






\section{Introduction}

The search for the holy grail in theoretical physics is composed of the main challenges that dwelt among us since the last century.  One of these challenges is to unify in a single and consistent framework both theories of quantum mechanics and general relativity.  The combination of special relativity and quantum field theory has already been accomplished through the Klein-Gordon and Dirac approaches.  However, the path to reconcile the general relativity with the quantum field theory is still a mystery.

This so-called quantization procedure of general relativity has stumbled onto another theoretical physics challenge, i.e., the infinities (divergences) that appear in some specific calculations during the quantization process.  This issue is directly connected to the understanding of the behavior of quantum fields at the high energy scale which is also connected to the structure of spacetime at (or near) the Planck scale.   To understand the structure of spacetime at this scale is necessary in order to construct the Hilbert space inner product, essential to the definition of the particle states.  There are several formalisms that deal with these questions and one of those is the noncommutative (NC) geometry, which can, for these reasons, be considered as a toy model for quantum gravity.

For example, one attempt to free us from the infinities that appear in quantum field theory was made by Snyder \cite{snyder} when he constructed a five dimensional NC algebra in order to define a minimum length for spacetime structure.  Unfortunately, a little time after the Snyder's effort, Yang \cite{yang} demonstrated that even in this Snyder's NC algebra, the divergences still persisted.

This result condemned Snyder noncommutativity (NVY) to be outcast for more than fifty years until Seiberg and Witten \cite{sw} demonstrated that the algebra resulting from string theory embedded in a magnetic field showed itself to have a NC algebra.  Since then we have seen an massive research production concerning several NC formulations that deserves our attention and investigation.

In a suitable basis, the algebra underlying Snyder's spacetime may be presented as a modification of the canonical commutation relations of phase-space, given by
\baa
\label{azero}
\[\hat{x}^{\mu}, \hat{x}^{\nu} \] &=& i\,\ell^2\,\hbar^{-1}\,\left(\hat{x}^{\mu}\,\hat{p}^{\nu}\,-\,\hat{x}^{\nu}\,\hat{p}^{\mu} \right)\,\,, \nonumber \\
\[\hat{x}^{\mu}, \hat{p}^{\nu} \] &=& i\,\hbar\,\eta^{\mu\nu}\,+\,i\,\ell^2\,\hbar^{-1}\,\hat{p}^{\mu}\,\hat{p}^{\nu} \,\,,\\
\[\hat{p}^{\mu}, \hat{p}^{\nu} \]&=&0 \; . \nonumber
\eaa
This algebra evolves a fundamental minimal length $\ell$, the scale of NCY, such that the classical phase space of quantum mechanics is recovered at
$\ell = 0$. The commutation relations (\ref{azero}) describe a discrete space-time which, at the same time, is compatible with Lorentz invariance. The original motivation behind these relations was that the introduction of the length scale $\ell$ is tantamount to regarding hadrons in quantum field theory as extended objects, because at the time renormalization theory was regarded as a distasteful procedure \cite{snyder,yang}. But, the success of the renormalization method resulted in little attention being paid to the subject for some time.

In the 1980's, NC geometry was considered as a way to extend the standard model in a number of different ways \cite{douglas,szabo}. In condensed matter physics, NCY appears naturally. For example, NC geometry describes the dynamics of electrons in a magnetic field at the lowest energy level which is related to the quantum Hall effect \cite{hall}. Recent interest in NC geometry is strongly motivated by the discovery that string theory leads to NC geometry in certain limits. In matrix models of M-theory, for example, compactification leads to NC tori; open strings in magnetic fields are described by NC algebra with the Moyal product. As we have mentioned above, the so-called SW map \cite{sw} between commutative and NC gauge theories explained that gauge symmetries, including diffeomorphisms, can be realized by standard commutative transformations on commutative fields.

This theoretical framework is called NC field theory and it may be a relevant physical model at scales in between $\ell_P \;(\simeq 1.6 \times 10^{-33} \mbox{cm})$ and $\ell_{LHC}\; (\simeq 2 \times 10^{-18} \mbox{cm})$. In fact, one of the main threads of research in this field has been related to studies of energetic cosmic rays, as we will discuss further below. In the following we will study this relationship in some detail. These field theories provide fruitful avenues of exploration for several reasons, that will be explained in more depth below.

Firstly, some quantum field theories are better behaved on NC space-time than on ordinary space-time. In fact, some are completely finite, even non-perturbatively. In this manner spacetime NCY presents an alternative to supersymmetry or string theory. Secondly, it is a useful arena for studying physics beyond the standard model, and also for standard physics in strong external fields. Thirdly, it sheds light on alternative lines of attack to address various fundamental issues in QFT, for instance the renormalization and axiomatic programs. Finally, it naturally relates field theory to gravity. Since the field theory may be easier to quantize, this may provide significant insights into the problem of quantizing gravity.
However, in his approach, Snyder postulated an identity between coordinates and generators of the $SO(4,1)$ algebra.   Hence, he promoted the spacetime coordinates to Hermitian operators.

Since to loose Lorentz invariance property is not a good thing for any theory, Doplicher, Fredenhagen and Roberts (DFR), have suggested that the NC parameter may not be a constant one and in this way the Lorentz invariance would be recovered.  We will see that the DFR algebra has been proposed based on issues that come from general relativity and quantum mechanics.
The authors claim that very accurate measurements of spacetime localization could transfer energies to test particles that at least theoretically could be
sufficient to create a gravitational field that in principle could trap photons.
But no mention at all was made by the authors concerning the fact that this non-constant parameter must have an associated momentum.   In a recent work \cite{mateus}, one of us has indicated that in fact if the NC parameter is a coordinate of this new Hilbert space, an associate momentum is directly connected to it.   Namely the new phase space would be formed by the original coordinates and the ($\theta,\pi$) new pair.   This new configuration brings new consequences in the so-called NC QFT and in this work we will analyze some of these consequences.   In this way we have to consider this formalism as an extension of the DFR one and because of that we have called it as DFR-extended or DFR$^*$.  Both will be used at random.   However, from the moment we believed that it was proved that DFR=DFR$^*$, we have used DFR to mean a $(x, p, \theta, \pi)$ NC $\theta$-variable phase-space.

We will follow a sequence that can provide the interested reader with the basics of this extension of the DFR formalism.  In this way, in the second section we have described some of the DFR$^*$ basic concepts.  In section III we have reviewed the case of the NC DFR$^*$ harmonic oscillator to help the understanding.  In section IV we have discussed the DFR-extended analysis of the NC relativistic particle in extra dimensions.   This model brought interesting results besides the ones connected directly to what we want to show.
In the section V we have computed the basic commutation relations for the scalar field in the DFR$^*$ phase-space and consequently, we have filled this gap and others in $\theta$-literature that do not see the existence of this associated momentum.   Finally, in section VI, the conclusions and perspectives were depicted.   We have constructed an Appendix with some explanations about Moyal-Weyl product.

\section{The basics of DFR-extended formalism}

One of the most popular approaches of NCY \cite{douglas,szabo} is the one governed by the well known Moyal-Weyl product.  In this approach, the standard product of two NC objects is substituted by the so-called star-product given by
\bbe
\label{A}
\hat{f}(x) \star \hat{g}(x)\,=\,\mbox{exp}\Big(\frac i2 \, \theta^{\mu\nu} \partial_{x\mu} \partial_{y\nu} \Big)\, \hat{f}(x) \, \hat{g}(y) \Big|_{x=y} \; ,
\eee
\noindent where  $\theta^{\mu\nu}$ is the well known NC parameter that is present at the beginning point of the NC definition, namely,
\bbe
\label{B}
\[ \hat{x}^{\mu}, \hat{x}^{\nu} \]\,=\,i\,\theta^{\mu\nu}\,\,,
\eee
\noindent where the hat notation indicates an operator as in Snyder's formalism.  The constant feature of $\theta^{\mu\nu}$ brings another problem, since we have a fix direction defined the right side of Eq. (\ref{B}).  This fact breaks the Lorentz invariance of the theory.  In the Appendix we have described the MW-product with more details.

In few words, one solution of this problem leads us to another formulation of a NC theory formulated by Doplicher, Fredenhagen and Roberts (DFR) \cite{dfr}, which is based in general relativity and quantum mechanics arguments.  This formalism recover Lorentz invariance through the promotion of $\theta^{\mu\nu}$ to be a standard coordinate of the system, that is, $\theta^{\mu\nu}$ in (\ref{B}) is promote to an operator $\hat{\theta}^{\mu\nu}$.  Of course, being the coordinate, the algebra turns out to be, together with Eq. (\ref{A})

\bbe
\label{C}
\[ \hat{x}^{\mu} , \hat{p}^{\nu} \]\,=\, i \eta^{\mu\nu} \; , \;
\[\hat{x}^{\mu} , \hat{\theta}^{\mu\nu} \]\,=\[\hat{p}^{\mu} , \hat{p}^{\nu}\]\,=\,\[ \hat{\theta}^{\mu\nu} , \hat{\theta}^{\rho\lambda}\]\,=\,0 \; ,
\eee

\noindent which completes the DFR algebra.


In this work, which in some way, could be considered as a continuation of a recent one \cite{mateus} by one of us with collaborators, we show that the momentum $\pi^{\mu\nu}$ is directly connected to Lorentz invariance of the theory.  Considering this scenario we will construct the complex scalar field theory and its physical consequences.   Although some QFT-DFR analysis was accomplished \cite{saxell}, it did not considered the existence and the necessity of a momentum conjugate to the coordinate $\theta^{\mu\nu}$.  Here, we developed this momentum operator algebra through the description of the scalar QFT.



In order to make this paper self-contained we will briefly talk about the main steps of the extended DFR (which from now on we will call DFR$^*$) algebra that contains also a momentum associated to $\theta^{\mu\nu}$ \cite{alexei,japoneses,amorim}.  We have quoted linear because this is a  momentum constructed in a $D=10\;(4+6)$ space-time, where four coordinates are Minkowskian and the other six compose the $\theta$-space which is unphysical.

The set of standard variables of a NC theory is given by $(x^{\mu},p^{\mu})$ which has also a NC parameters, $\theta^{\mu\nu}$, introduced in the theory through Eq. (\ref{A}).  The DFR theory was already exposed in section I.  It has also $\theta^{\mu\nu}$ but now this last one is a usual variable of the system and the DFR-algebra was depicted en Eq. (\ref{C}).  So, the DFR$^*$ phase-space is $(x^{\mu},p^{\mu},\theta^{\mu\nu},\pi^{\mu\nu})$.
In the next section we will show \cite{mateus} that this is the correct phase-space for the DFR$^*$ formalism, since the existence of $\pi^{\mu\nu}$ is connected to Lorentz invariance \cite{ccz}.

This so-called DFR$^*$ algebra can be described, combined with the ones given in Eq. (\ref{A}) nd (\ref{C}), by
\bbe
\label{D}
\[\hat{p}^{\mu},\hat{\pi}^{\nu\lambda} \]\,=\,0 \quad,\quad\[\hat{\theta}^{\mu\nu},\hat{\pi}^{\rho\lambda} \]\,=\,i\, \uma^{\mu\nu\rho\lambda} \quad,\quad\[\hat{x}^{\mu},\hat{\pi}^{\nu\lambda} \]\,=\, \frac i2 \, \uma^{\mu\rho\nu\lambda} \,\hat{p}_{\rho}\,\,,
\eee
\noindent where $\uma^{\mu\nu\rho\lambda}:=\eta^{\mu\rho}\,\eta^{\nu\lambda}\,-\,\eta^{\mu\lambda}\,\eta^{\nu\rho}$, and some elements described above can form the Jacobi identity
\bbe
\label{E}
[[\hat{x}^{\mu},\hat{\pi}^{\nu\lambda}],\hat{x}^{\rho}]\,+\,[[\hat{x}^{\rho},\hat{x}^{\mu}],\hat{\pi}^{\nu\lambda}]\,
+\,[[\hat{\pi}^{\nu\lambda},\hat{x}^{\rho}],\hat{x}^{\mu}]\,=\,0 \; ,
\eee
\noindent where, using (\ref{D}) results in
\bbe
\label{F}
[[\hat{x}^{\mu},\hat{\pi}^{\nu\lambda}],\hat{x}^{\rho}]\,-\,[[\hat{x}^{\rho},\hat{\pi}^{\nu\lambda}],\hat{x}^{\mu}]\,=\,-\,\uma^{\mu\rho\nu\lambda} \; ,
\eee
\noindent and this completes the DFR-extended algebra. DFR \cite{dfr} demonstrated that a state that minimizes the uncertainty in space and time in a specific Lorentz frame is tantamount to integrate the tensor $\theta^{\mu\nu}$ over spatial rotations.   This formalisms leads us to a rotational but not a Lorentz invariant theory.  In operator terms, we can say that with the NC parameter operator we can obtain the canonical case after choosing an eigenstate of the operator.  In this case, the commutator $[\hat{x}^{\mu},\hat{x}^{\nu}]$ is equal to an eigenvalue.   CCZ \cite{ccz}, in order to have an entire Lorentz invariance, developed a NC field theory with integration over all values of $\theta$. The interested reader can see more details in \cite{sigma} and references therein.

\section{The DFR-extended harmonic oscillator}

In \cite{mateus} the authors have analyzed  an harmonic oscillator constructed in a DFR-extended \cite{amorim} phase-space.  The generalized Hamiltonian is given by

\bbe
\label{G}
H\,=\frac{p_{i}^2}{2m}\,+\,\frac{\pi_{ij}^2}{2\Lambda}\,+\,V(x_i,p_i,\theta_{ij},\pi_{ij})\,\,,
\eee

\noindent where $\Lambda$ is a parameter with $(\mbox{length})^{-3}$ dimension and the potential $V$ is a function of DFR-extended variables.
Let us define the following symplectic variables $\xi^i$ as being $(x^i,\,p_i,\,\theta^{ij},\,\pi_{ij})$.
We can write the generalized Poisson bracket for this system in a compact and symplectic form as
\baa
\label{gpb0}
\left\{ F,G\right\}   =  \left\{ \xi^{i},\xi^{j}\right\} \frac{\partial F}{\partial\xi^{i}}\frac{\partial G}{\partial\xi^{j}} \; ,
\eaa
\ni where we are using the sum rule for repeated indices. Hence, following (\ref{gpb0}), the equations of motion for
$(x^i,\,p_i ,\,\theta^{ij},\, \pi_{ij})$ are given by
\bbe
\label{H1}
\dot{x}^i \,=\, \,\frac{p^i}{m} \, + \, \frac{\partial V}{\partial p^i} \, + \, \theta^{ij} \frac{\partial V}{\partial x^j}\,+\,\Big( \frac{\pi^{ji}}{\Lambda}\,+\,\frac{\partial V}{\partial \pi_{ji}} \Big) \,p_j
\eee

\bbe
\label{H2}
\hspace{-6.1cm}\dot{p}_i\,=\,-\frac{\partial\,V}{\partial\,x^i}
\eee

\bbe
\label{H3}
\hspace{-1cm}\dot{\theta}^{ij}\,=\,\frac 2\Lambda \pi^{ij}\,+\,2\frac{\partial\,V}{\partial\,\pi_{ij}}
\eee

\bbe
\label{H4}
\dot{\pi}_{ij}\,=\,-\,2\,\frac{\partial\,V}{\partial\,\theta^{ij}}\,+\,\frac{\partial\,V}{\partial\,x^{i}}\,p_j \; ,
\eee
\noindent and it can be seen clearly that when $\pi_{ij}=0$ (namely, when the phase-space is $(x,p, \theta))$ the first consequence is that the potential $V$ will not be a function of $\pi_{ij}$ and to construct Eq. (\ref{H4}) makes no sense.  The second consequence is that, from Eq. (\ref{H3}), when $\pi_{ij}=0$ we have that $\theta^{ij}=\mbox{constant}$, and therefore the Lorentz invariance is lost and we have a canonical NCY.  Let us continue with a specific construction for the potential $V$, for example.

In \cite{amorim} an isotropic NCHO was constructed in a $D=9$ DFR-extended phase-space.  The extended potential was given by

\bbe
\label{I}
V(x^i , p_i , \theta^{ij},\pi_{ij})\,=\,\frac 12 m \omega^2 \Big( x^i\,+\, \frac 12 \theta^{ij} p_j\Big)^{2}\,+\, \frac 12 \Lambda \Omega^2 \theta^2 \; ,
\eee

\noindent and he extended Hamiltonian can be written as

\bbe
\label{J}
H\,= \, \frac{p_{i}^{2}}{2m} \, + \,\frac{\pi_{ij}^{2}}{2\Lambda} \,+\, \frac 12 m \omega^2 \Big( x^i\,+\,\frac 12 \theta^{ij} p_j \Big)^{2} \,+\, \frac 12 \Lambda \Omega^2 \theta^2\,\,.
\eee

\noindent Consequently, the equations of motion are

\bbe
\label{J1}
\dot{x}^i \,=
\, \frac{p^i}{m} \, + \, \frac 12 \, \theta^{ij} \Big( m \omega^2 x_j \,
+\, \frac 12 m \omega^2 \theta_{jl} p^l \Big)\,+\,\frac{\pi^{ij}}{\Lambda}\,p_j\,\,,
\eee

\bbe
\label{J2}
\!\!\!\!\!\!\!\!\!\!\!\!\!\!\!\!\!\!\!\!\!\!\!\!\!\!\!\!\!\!\!\!\!\!\!\!\!\!\!\!\!\!\!\!\!\!\!\!\!\!\dot{p}_i\,=\,-\,m\omega^2 x_i\,-\,\frac 12 m\omega^2 \theta_{ij}\,p^j\,\,,
\eee

\bbe
\label{J3}
\!\!\!\!\!\!\!\!\!\!\!\!\!\!\!\!\!\!\!\!\!\!\!\!\!\!\!\!\!\!\!\!\!\!\!\!\!\!\!\!\!\!\!\!\!\!\!\!\!\!\!\!\!\!\!\!\!\!\!\!\!\!\!\!\!\!\!\!\!\!\!\!\!\!\!\!\!\!\!\!\!\!\!\!\!\!\!\!\!\!\!\!\!\!\!\!\!\dot{\theta}^{ij}\,=\,\frac 2\Lambda \, \pi^{ij}\,\,,
\eee

\bbe
\label{J4}
\!\!\!\!\!\!\!\!\!\!\!\!\!\!\!\!\!\!\!\!\!\!\!\!\!\!\!\!\!\!\!\!\!\!\!\!\!\!\!\!\!\!\!\!\!\!\!\!\!\!\!\!\!\!\!\!\!\!\!\!\!\!\!\!\!\!\!\!\!\!\!\!\!\!\!\!\!\!\!\!\!\!\!\dot{\pi}_{ij}\,=\,-\,2\,\Lambda \Omega^2 \theta_{ij}\,\,.
\eee

In a naive way, it would be possible that we could understand that when $\pi_{ij}$ = 0 it would be easy to conclude
that the resulting phase-space would be given by the DFR one. However, as we mentioned
before when we have analyzed the equations of motion for $\theta^{ij}$ and $\pi_{ij}$ in Eqs. (\ref{H3}) and
(\ref{H4}) respectively, we can see that $\theta_{ij} = \mbox{constant}$. If $\pi_{ij}$ = 0 in (\ref{J3}) we can see
clearly that $\theta^{ij} = \mbox{constant}$. If we construct a Hamiltonian independent of $\pi_{ij}$ it does not make
sense to construct Eqs. (\ref{H4}) and (\ref{J4}). Substituting these values in Eqs. (\ref{J1}) and (\ref{J2})
we recover the canonical NCY and not the DFR NCY approach.
Consequently we can conclude that the DFR-extended and pure DFR formalisms are
both connected to the canonical NCY via $\pi_{ij}$ and not only via the nature of $\theta^{ij}$. Namely,
to carry out a dimensional reduction of the phase-space (doing $\pi_{ij} = 0$) means that $\theta^{ij}$ loses automatically
its variable parameter identity and becomes again a constant parameter. Hence,
the phase-space dimensional reduction would be represented by $(x^i,p_i,\theta^{ij},\pi_{ij}) \longrightarrow (x^i,p_i)$ where
$\theta^{ij}$ is only a constant
parameter, the result of the operatorial bracket between $x$'s.  The Lorentz invariance is lost and the NCY is the canonical one.

So, concerning the original DFR formalism, although in general, the momentum $\pi_{ij}$
may not be relevant, we understand that the momentum associated to $\theta^{ij}$ is necessary. As
a matter of fact, it would be natural and direct to construct this object since $\theta^{ij}$, in DFR
phase-space, is a coordinate and must have an associated momentum. However, what is
new, in our point of view, is to connect the existence of $\pi_{ij}$ with the kind of the NCY or,
in other words, if the NCY is DFR-extended or canonical.

This result make us think that, if we consider, for example, QFT's
systems embedded in a NC space-time, the implications are even more serious because the
existence of a $\theta^{\mu\nu}$-variable NC parameter recovers the Lorentz invariance of the NC theory.
But, the relevance of $\pi_{\mu\nu} = 0$ is the fact that it brings back a constant $\theta^{\mu\nu}$, and hence
we have the Lorentz invariance violated. So, the connection between
both objects ($\theta^{\mu\nu}$ and $\pi_{\mu\nu}$) is a connection between Lorentz invariant or non-invariant NC
theories. Besides, we will see that the momenta $\pi_{\mu\nu}$ allows us to construct the commutation relations for the scalar field in DFR phase-space.

Back to Eqs. (\ref{J1})-(\ref{J4}) we can see that, in this specific example that, from Eq. (\ref{J3}), if $\theta=\,\mbox{constant}$ $\Longrightarrow \pi=0$ and Eq. (\ref{J4}) makes no sense at all.  Hence, we have the inverse condition, i.e., $\theta=\,\mbox{constant}$. $\Longrightarrow \pi=0$, which is the inverse of $\pi=0 \Longrightarrow \theta =\,\mbox{constant}$.  Let us see another example, the NC relativistic particle to reinforce these claims above.

\section{The NC relativistic particle}

In \cite{alexei}, the author proposed that the cure for the lack of relativistic invariance for NC models is to modify the constant feature of the NC parameter.   Consequently, he has analyzed the NC version for D-dimensional relativistic particle with a $\theta$-variable phase-space and a $\pi$-momentum.

Since we are interested in the dynamics of the phase-space, we have calculated the equations of motion and the NC relativistic acceleration in order to discuss the $\theta_{constant} \Longrightarrow \theta_{variable}$ duality and its consequence.  We will see that although the algebra is not the DFR one the consequences of the duality are kept, namely, if we have a $\theta$-variable the phase-space must have the $\pi$-momentum (the DFR-momentum).

\subsection{Noncommutative relativistic free particle}

In this section, since we are interesting in the DFR features that exist in the analyzed model, we will mention only the relative points of \cite{alexei} where more details can be found.

The Lagrangian of NC free relativistic particle is

\bbe
S(x,\theta)=\int d\tau\left[\dot{x}^{\mu}v_{\mu}-\frac{e}{2}\left(v^{2}-m^{2}\right)+\frac{1}{\theta^{2}}\dot{v}_{\mu}\theta^{\mu\nu}v_{\nu}\right] \; ,
\eee
\ni where $\theta^{2}\equiv\theta^{\mu\nu}\theta_{\mu\nu}$, $\eta=\mbox{diag}(+,-,...,-)$.
And $p^{\mu},\,\pi^{\mu},\,p_{e},\,p_{\theta}^{\mu\nu}$ are the conjugate momenta
associated to $x^{\mu}(\tau),\,v^{\mu}(\tau),\,e(\tau)$ and $\theta^{\mu\nu}(\tau)$, respectively. We will use the fundamental algebra \cite{alexei} defined  by
\bbe
\label{algebra}
\left\{ x^{\mu},x^{\nu}\right\} =-\frac{2}{\theta^2}\,\theta^{\mu\nu},\qquad \left\{ x^{\mu},p^{\nu}\right\} =\eta^{\mu\nu},\qquad  \left\{ v^{\mu},\pi^{\nu}\right\} =\eta^{\mu\nu} \; ,
\eee
\bbe  \left\{ x^{\mu},v^{\nu}\right\} =\eta^{\mu\nu} \qquad  \left\{ x^{\mu},\pi^{\nu}\right\} =-\frac{1}{\theta^2} \, \theta^{\mu\nu}\eee
\bbe \left\{ \theta_{\mu\nu},p_{\theta}^{\rho\sigma}\right\} = - \delta^{[\rho}_{\mu}\,\delta^{\sigma ]}_{\nu}\eee
\bbe \left\{ x^{\mu},p_{\theta}^{\rho\sigma}\right\} = -\left\{ \pi^{\mu},p_{\theta}^{\rho\sigma}\right\} =
\frac{1}{\theta^2}\,\eta^{\nu[\rho}v^{\sigma]}\,-\,\frac{4}{\theta^4}(\theta v)^{\mu} \theta^{\rho\sigma} \; .
\eee
This system is singular and has the following primary constraints
\baa
G^{\mu}  &=&  p^{\mu}-v^{\mu}\\
T^{\mu}  &=&  \pi^{\mu}-\frac{1}{\theta^{2}}\theta^{\mu\nu}v_{\nu}\\
p_{\theta}^{\mu\nu}  &=&  0\\
p_{e} & =&  0 \; ,
\eaa

\ni and we can write the total Hamiltonian as being

\bbe
H=\frac{e}{2}\left(v^{2}-m^{2}\right)+\lambda_{1\mu}G^{\mu}+\lambda_{2\mu}T^{\mu}+\lambda_{e}p_{e}+\lambda_{\theta\mu\nu}p_{\theta}^{\mu\nu} \; ,
\eee

\ni where the $\lambda$'s are the Lagrange multipliers. Using the time consistency we have the secondary constraint

\bbe
K\equiv v^{2}-m^{2}=0 \; ,
\eee

\ni and other relations that allow us to determine the  Lagrange multipliers

\baa
\dot{G^{\mu}}=\left\{ G^{\mu},H\right\} =0 & \Longrightarrow & \lambda_{2}^{\mu}=0\\
\dot{T}^{\mu}=\left\{ T^{\mu},H\right\} =0 & \Longrightarrow & \lambda_{1}^{\mu}=ev^{\mu}+\frac{2}{\theta^{2}}\left(\lambda_{\theta}v\right)^{\mu}-\frac{4}{\theta^{4}}\left(\theta\lambda_{\theta}\right)\left(\theta v\right)^{\mu}\,\, .
\eaa
If we substitute the fixed Lagrange multipliers into the Hamiltonian
we have that
\bbe
\label{AAA}
H=\frac{e}{2}\left(p^{2}-m^{2}\right)+\left(ev_{\mu}+\frac{2}{\theta^{2}}\left(\lambda_{\theta}v\right)_{\mu}-\frac{4}{\theta^{4}}\left(\theta\lambda_{\theta}\right)\left(\theta v\right)_{\mu}\right)\left(p^{\mu}-v^{\mu}\right)+\lambda_{e}p_{e}+\lambda_{\theta\mu\nu}p_{\theta}^{\mu\nu} \; ,
\eee

\ni and it can be seen we were left with two undetermined Lagrange multipliers.

In the same way as we have carried out to construct Eq. (\ref{gpb0}) we will define the following symplectic variables
\baa
\label{AAb}
\xi^{\mu} & \equiv & \left(x^{\mu},p_{\mu}\right) \nonumber \\
\zeta^{\mu} & \equiv & \left(v^{\mu},\pi_{\mu}\right) \nonumber \\
\chi^{\mu} & \equiv & \left(e,p_{e}\right) \nonumber \\
\Omega^{\mu\nu} & \equiv & \left(\theta^{\mu\nu},p_{\theta\mu\nu}\right) \; .
\eaa
We can write the Poisson brackets for this system in a compact and symplectic form
as follow
\baa
\label{gpb}
\left\{ F,G\right\}  & = & \left\{ \xi^{\mu},\xi^{\nu}\right\} \frac{\partial F}{\partial\xi^{\mu}}\frac{\partial G}{\partial\xi^{\nu}}+\left\{ \zeta^{\mu},\zeta^{\nu}\right\} \frac{\partial F}{\partial\zeta^{\mu}}\frac{\partial G}{\partial\zeta^{\nu}} \nonumber \\
 & + & \left\{ \chi^{\mu},\chi^{\nu}\right\} \frac{\partial F}{\partial\chi^{\mu}}\frac{\partial G}{\partial\chi^{\nu}}+\left\{ \Omega^{\mu\nu},\Omega^{\rho\sigma}\right\} \frac{\partial F}{\partial\Omega^{\mu\nu}}\frac{\partial G}{\partial\Omega^{\rho\sigma}}\,\,.
\eaa
According to the (\ref{gpb}) we obtain the following equation of motion for $x^{\mu}$ and $p^{\mu}$
\baa
\label{AAc}
\dot{x}^{\mu}&=&\left\{ x^{\mu},H\right\} \nonumber
\\
&=&\left\{ x^{\alpha},p_{\beta}\right\} \frac{\partial x^{\mu}}{\partial x^{\alpha}}\frac{\partial H}{\partial p_{\beta}}+\left\{ p_{\beta},x^{\alpha}\right\} \frac{\partial x^{\mu}}{\partial p_{\beta}}\frac{\partial H}{\partial x^{\alpha}}+\left\{ x^{\alpha},x^{\beta}\right\} \frac{\partial x^{\mu}}{\partial x^{\alpha}}\frac{\partial H}{\partial x^{\beta}}\nonumber \\
&+&\left\{ x^{\alpha},p_{\theta\rho\sigma}\right\} \frac{\partial x^{\mu}}{\partial x^{\alpha}}\frac{\partial H}{\partial p_{\theta\rho\sigma}}+\left\{ p_{\theta\rho\sigma},x^{\alpha}\right\} \frac{\partial x^{\mu}}{\partial p_{\theta\rho\sigma}}\frac{\partial H}{\partial x^{\alpha}}
\eaa
\bbe
\label{AAD}
\Rightarrow\dot{x}^{\mu}=ep^{\mu}+\frac{2}{\theta^{2}}\left(\lambda_{\theta}v\right)^{\mu}-\frac{4}{\theta^{4}}\left(\theta\lambda_{\theta}\right)\left(\theta v\right)^{\mu} \; ,
\eee

\ni and for $p_{\mu}$ we have that

\baa
\label{AAe}
\dot{p}^{\mu}&=&\left\{ p^{\mu},H\right\} \nonumber \\
&=&\left\{ x^{\alpha},p_{\beta}\right\} \frac{\partial p^{\mu}}{\partial x^{\alpha}}\frac{\partial H}{\partial p_{\beta}}+\left\{ p_{\beta},x^{\alpha}\right\} \frac{\partial p^{\mu}}{\partial p_{\beta}}\frac{\partial H}{\partial x^{\alpha}} \Rightarrow\dot{p}^{\mu}=0 \; .
\eaa
Analogously, we can compute the equations of motion for the other variables, namely,
\bbe
\label{AAB}
\dot{\theta}^{\mu\nu}\,=\,-\,2\,\lambda_{\theta}^{\mu\nu}
\eee
\bbe
\label{AAB2}
\dot{v}_{\mu}\,=\,0
\eee
\bbe
\label{AAB3}
\dot{e}\,=\,\lambda_e
\eee
\bbe
\label{AAB4}
\dot{p}_e\,=\,-\,v\cdot p\,+\,\frac 12 (v^2\,+\,m^2)
\eee
\bbe
\label{AAB5}
\dot{\pi}^{\mu}\,=\,\frac{4}{\theta^4} (\theta \lambda_{\theta})\,(\theta\,v)^{\mu}\,-\,\frac{1}{\theta^2}\eta^{\mu[\rho} v^{\sigma]}\,\lambda_{\theta\rho\sigma}
\eee
\bbe
\label{AAC}
\dot{p}^{\mu\nu}_{\theta}\,=\,\frac{8}{\theta^4}\,\left[ \frac{\theta^{\mu\nu}}{\theta^2} (\theta \lambda_{\theta}) (\theta v)^{\sigma} p_{\sigma}\,-\,\lambda_{\theta}^{\mu\nu} (\theta v)^{\sigma} p_{\sigma}\,-\,\theta^{\mu\nu} (\lambda_{\theta} v)^{\sigma} p_{\sigma}\,+\, \frac 12 (\theta \lambda_{\theta}) v^{[\mu} p^{\nu]} \right]\,\,.
\eee

Finally, in the same way we can calculate the acceleration in this NC phase-space, namely, $\ddot{x}^{\mu}\,=\,\{\dot{x},\,H \}$, which brings us the result

\bbe
\label{AAE}
\ddot{x}\,=\,\frac{8}{\theta^4} \left[ (\theta \lambda_{\theta})\,(\lambda_{\theta} v)^{\mu}\,-\,\frac{4}{\theta^2}\,(\theta \lambda_{\theta})^2\,(\theta v)^{\mu}\,+\,\lambda^2 (\theta v)^{\mu}\,-\,(\theta \lambda_{\theta}) (\lambda_{\theta} v)^{\mu} \right] \; ,
\eee

\ni where $\lambda^2\,:=\,\lambda_{\theta\mu\nu} \lambda_{\theta}^{\mu\nu}$.  This last result is very interesting since  the equation of motion (\ref{AAB}) shows us that if we have that $\theta=\mbox{constant}$, we have that $\lambda_{\theta}=0$.   In this way we will not have $p_{\theta}$ in the Hamiltonian written in (\ref{AAA}).   However, we can easily see from Eq. (\ref{AAC}) that we have that $\lambda_{\theta}=0 \Longrightarrow \dot{p}_{\theta}=0 \Longrightarrow p_{\theta}=\mbox{constant}$, but the important fact is that the phase-space for the Hamiltonian in Eq. (\ref{AAA}) will not have $p_{\theta}$.  Hence, although the algebra in Eq. (\ref{algebra}) is not a DFR$^*$ one, the scenario is the same, namely, if $\theta$ is not constant, the NC phase-space contains $p_{\theta}$, if $\theta$ is constant, we do not have $p_{\theta}$ within the phase-space. Notice that although the $\lambda$'s are auxiliary variables in order to construct the total Hamiltonian, they are connected to the momenta, by construction of the constraints formalism.

We can also notice that if $\theta = \mbox{constant}$ in Eq. (\ref{AAE}), the acceleration is zero,  This is an interesting result since we do not have any time derivative of $\theta$ in Eq. (\ref{AAE}) but this result is a consequence of the zeroness of $\lambda_{\theta}$.  However, the time derivative of $x^{\mu}$ in Eq. (\ref{AAD}) is not zero when $\lambda_{\theta} =0$ neither it is constant since $e(\tau)$ is variable ($p_{\mu}$ is constant since $\dot{p}_{\mu}=0$).


\section{Quantum NC scalar field theory}

In this section we will construct the first basic step of a QFT with the phase-space definitions established in the last section.  Since we have shown that the DFR and DFR-extended phase-space are in fact the same, we will use the name DFR to define the formalism embedded in the complete phase-space $(x,p,\theta,\pi)$.

In the papers developed by two of us \cite{an1,an2,an3}, one can see that the construction of the commutation relations between the bosonic/fermionic fields with themselves and with its associated momenta are missing.  It is our intention in this section to fill this gap.  In other words, we will demonstrate precisely the basic commutation relations using only the DFR elements.  The fermionic construction is an ongoing research that will be published in a near future.

In other papers that considers the DFR formalism or a kind of, such as \cite{dfr,ccz,he,ck,eh,saxell,alexei,japoneses} for example, we can find this basic step in an indirect way where the associated momenta are not defined.  The quantity used to construct the scalar field, that was used to be associated with the variable $\theta$, is a scalar quantity with no definition at all.  The consequence
Now we know that this last object is in fact the momenta associated with the NC parameter and this fact allows us to work with a well defined phase-space, the DFR one.

After the considerations given above, we can complement (clarify) \cite{ccz,saxell} by constructing the Fourier transform, so we can write a map between a member of the operator algebra and an ordinary function

\bbe
\hat{f}(\hat{x},\hat{\theta})\,=\,\int \frac{d^4 p}{(2\pi)^4}\frac{d^6 \pi}{(2\pi)^6}\, \, e^{-i(p\cdot \hat{x}\,+\,\pi\cdot \hat{\theta})}\,\widetilde{f} (p,\pi) \; ,
\eee

\ni where $\widetilde{f}$ is the Fourier transform

\bbe
\widetilde{f} (p,\pi)=\,\int d^4x \, d^6\theta \, \, e^{i(p\cdot x\,+\,\pi\cdot \theta)}\, f (x,\theta) \; ,
\eee

\ni where $p \cdot \hat{x} = p_{\mu} \hat{x}^{\mu}$ and $\pi \cdot \hat{\theta} = \frac 12 \, \pi_{\mu\nu} \hat{\theta}^{\mu\nu}$ ( the $1/2$ factor avoids the sum over repeated terms).  And the integration measures are
\baa
d^6 \pi \,&=&\, d \pi_{01} \,d \pi_{02}\, d \pi_{03} \,d \pi_{12}\, d \pi_{13}\, d \pi_{23} \nonumber \\
\mbox{} \nonumber \\
d^6 \theta &=& d \theta^{01} d \theta^{02} d \theta^{03} d \theta^{12} d \theta^{13} d \theta^{23} \; .
\eaa
\ni The details about $\theta$ and $\pi$ are described in \cite{amorim}.   But notice that in \cite{amorim} (and references therein), $\theta$-variable and $\pi$ are not necessarily connected as we have discussed so far.   It is important to say that we have clarified the one other main points treated in \cite{ccz} and \cite{saxell}.   In these last ones, the objects were described with a not well defined quantity coupled to $\theta^{\mu\nu}$;   Here we have demonstrated precisely that this quantity is the momentum $\pi$ which completes the DFR phase-space.

Since Eqs. (\ref{B}), (\ref{C}) and (\ref{D}) closes the extended DFR algebra, we can use the fact that the momentum $\pi_{\mu\nu}$ makes part
of the NC phase-space, let us construct the operator field in this DFR-extended algebra in Weyl representation \cite{saxell}

\bbe
\label{AA}
\hat{\phi} (\hat{x},\hat{\theta} )\,=\, \int \frac{d^4p}{(2\pi)^4} \, \frac{d^6 \pi }{(2\pi)^{6}} \, \, \widetilde{\phi} (p, \pi) \, \,
 e^{i(p \cdot\hat{x}+\pi\cdot \hat{\theta} )} \; ,
\eee

\noindent where $\widetilde{\phi}(p,\pi )$ is the Fourier transform of $\hat{\phi}(\hat{x},\hat{\theta})$ and $d^6 \pi$ is a Lorentz invariant measure given above.  Notice that the difference between the issues explored here and in \cite{saxell} is that now we know that the phase-space is described by $(x,p,\theta,\pi)$.  In order to write the components of $\hat{\theta}^{\mu\nu}$, let us make the diagonalization operation \cite{saxell}

\bbe
\label{BB}
 \langle\theta |\,\hat{\phi} (\hat{x},\hat{\theta} )\, | \theta \rangle \,=\, \langle \theta |\,\int \frac{d^4 p}{(2\pi)^4} \, \frac{d^6 \pi}{(2\pi)^{6}} \,\widetilde{\phi} (p, \pi) \, \, e^{i(p \cdot\hat{x}+\pi\cdot \hat{\theta} )} \,|\theta \rangle \; ,
\eee

\noindent and the equal-time commutation relations for the operators $\hat{\phi} (\hat{x}, \hat{\theta}, t)$ and $\hat{\pi} (\hat{x}, \hat{\theta}, t)$ are

\bbe
\label{CC2}
\[\hat{\phi} (x,\theta, t),\hat{\phi} (x',\theta',t) \]\,=\,\[ \hat{\pi}(x,\theta,t),\hat{\pi}(x',\theta',t) \]\,=\,0 \; ,
\eee

\ni and

\baa
\label{CC1}
\[ \hat{\phi} (x,\theta,t),\hat{\pi}(x',\theta',t) \]\,=\, i\delta^3 ({\bf x}-{\bf x}^{\prime})\,\delta^6(\theta-\theta^{\prime}) \; ,
\eaa

\noindent where from now on, when we consider a $D=3$ phase-space, we are considering ${\bf x}$ as a vector. One can ask, since we have chosen the ordinary commutator, if the field quanta will obey a kind of Bose-Einstein statistics in this NC phase-space.

In \cite{saxell} the author has written an incomplete $\hat{\phi}(x,\theta)$ using the Weyl representation.  We say incomplete because now we know that $\theta^{\mu\nu}$ has an associated momentum given by $\pi_{\mu\nu}$.  In this way now we can expand the field operator $\hat{\phi} (x,\theta,t)$ with respect to a basis.  Let us use the set of plane waves such as

\bbe
\label{K}
u_{p,\pi} (x,\theta)\,=\,N_{p,\pi} \, \, e^{i(p\cdot x+\pi\cdot \theta )} \; ,
\eee

\noindent which means that we can write the Fourier modes as

\bbe
\label{LL}
\hat{\phi}(x,\theta,t)\,=\,\int d^3 {\bf p} \, d^6 \pi \, \, N_{p,\pi} \, \, e^{i(p\cdot x+\pi\cdot \theta)} \hat{a}_{p.\pi} (t) \ ,
\eee

\noindent where $N_{p,\pi}$ is a normalization constant.  If we substitute Eq. (\ref{LL}) into (\ref{JJ}) we will have that

\bbe
\label{MM}
\ddot{\hat{a}}_{p,\pi} (t)\,=\,-\, \Big({\bf p}^{2}\,+\,\frac{\lambda^2}{2} \, \pi^2\,+\,m^2 \Big) \,  \hat{a}_{p,\pi} (t) \; ,
\eee

\noindent which has a general solution given by

\bbe
\label{NN}
\hat{a}_{p,\pi} (t) \,=\, \hat{a}_{p,\pi}^{(1)} \, \, e^{-i\omega_{p,\pi} t} \,+\, \hat{a}_{p,\pi}^{(2)} \, \, e^{i\omega_{p,\pi} t} \; ,
\eee

\noindent in which the dispersion relation is

\bbe
\label{OO}
\omega_{p,\pi} \,=\, \sqrt{{\bf p}^{\,2}\,+\,\frac{\lambda^2}{2} \, \pi^2 \,+\,m^2} \; ,
\eee

\noindent and from (\ref{NN}) we can easily see that $\hat{a}_{p,\pi}^{(1)}$ and $\hat{a}_{p,\pi}^{(2)}$ are constants in time.  The real-valued feature of the classical field shows us that, of course, the operator is hermitian, hence,

\bbe
\label{PP}
(\hat{a}_{p,\pi}^{(1)} )^{\dagger}\,=\, \hat{a}_{-p,-\pi}^{(2)} \; ,
\eee

\noindent which is a standard constraint.  For now, we will associate $a_{p,\pi}$ and $a^{\dagger}_{p,\pi}$ with annihilation and creation operators, respectively, in DFR$^*$ formalism. Therefore, the basis expansion can be written as,

\bbe
\label{QQ}
\hat{\phi} (x,\theta,t)\,=\,\int d^3 {\bf p} \, d^6 \pi\,N_{p,\pi}\,\Big[ \hat{a}_{p,\pi} \, e^{i(p\cdot x+\pi\cdot \theta-\omega_{p,\pi} t)}\,+\,\hat{a}_{p,\pi}^{\dagger} \, e^{-i(p\cdot x+\pi\cdot \theta-\omega_{p,\pi} t)} \Big] \; ,
\eee

\noindent and we will see in a moment that the basis expansion of the conjugate field is given by
$\hat{\pi}(x,\theta)\,=\,\dot{\hat{\phi}}(x,\theta)$, so we have that

\bbe
\label{RR}
\hat{\pi}(x,\theta,t) \,=\,\int d^3 {\bf p} \, d^6 \pi\,N_{p,\pi}\,(-i\omega_{p,\pi} )\Big[ \hat{a}_{p,\pi}e^{i(p\cdot x+\pi\cdot \theta-\omega_{p,\pi} t)}\,-\,\hat{a}_{p,\pi}^{\dagger}e^{-i(p\cdot x+\pi\cdot \theta-\omega_{p,\pi} t)} \Big] \; ,
\eee

\noindent since $\hat{\pi}=\dot{\hat{\phi}}$. The free field can be expanded in terms of creation and annihilation operators, namely,

\bbe
\label{SS1}
\Big[\hat{a}_{p,\pi},\hat{a}_{p',\pi'}^{\dagger}\Big]\,=\,\delta^3 ({\bf p}-{\bf p}^{\prime})\,\delta^6 (\pi-\pi')
\eee
\bbe
\label{SS2}
\Big[\hat{a}_{p,\pi},\hat{a}_{p',\pi'}\Big]\,=\,\Big[\hat{a}_{p,\pi}^{\dagger},\hat{a}_{p',\pi'}^{\dagger}\Big]\,=\,0 \; .
\eee
Substituting Eq. (\ref{QQ}) in Eq. (\ref{CC2}) and using relations (\ref{SS1}) and (\ref{SS2}) we can construct the commutation relation given by
\baa
\label{TTzero}
&&\Big[ \hat{\phi}(x,\theta,t),\hat{\phi}(x',\theta', t) \Big]=\int d^{\,9}P \int d^{\,9}P' N_{p,\pi}\,N_{p',\pi'}\, \times  \nonumber \\
&\times& \bigglb\{\Big[\hat{a}_{p,\pi},\hat{a}_{p',\pi'}\Big]\,e^{i(p\cdot x+\pi\cdot \theta+p\cdot x'+\pi\cdot \theta')}
\,+\,\Big[\hat{a}_{p,\pi},\hat{a}_{p',\pi'}^{\dagger}\Big]\,e^{i(p\cdot x+\pi\cdot \theta-p\cdot x'-\pi\cdot \theta')} \nonumber \\
&& \,+\,\Big[\hat{a}_{p,\pi}^{\dagger},\hat{a}_{p',\pi'}\Big]\,e^{-i(p\cdot x+\pi\cdot \theta-p\cdot x'-\pi\cdot \theta')}
\,+\,\Big[\hat{a}_{p,\pi}^{\dagger},\hat{a}_{p',\pi'}^{\dagger}\Big]\,e^{-i(p\cdot x+\pi\cdot \theta+p\cdot x'+\pi\cdot \theta')} \biggrb\}=0 \; ,
\eaa

\noindent where $d^9 P = d^3 {\bf p} \,\, d^6 \pi$. Substituting the Eqs. (\ref{QQ}) and (\ref{RR}) into commutation relation (\ref{CC1})  we have that
\baa
\label{TT}
&&\Big[ \hat{\phi}(x,\theta,t),\hat{\pi}(x',\theta', t) \Big] \, =\, \int d^{\,9}P \int d^{\,9}P' \,  N_{p,\pi}\,N_{p',\pi'}\,(-i\omega_{p,\pi})\, \times
\nonumber \\
&\times& \bigglb\{\Big[\hat{a}_{p,\pi},\hat{a}_{p',\pi'}\Big]\,e^{-i(p\cdot x+\pi\cdot \theta-p\cdot x'-\pi\cdot \theta')}
\,-\,\Big[\hat{a}_{p,\pi},\hat{a}_{p',\pi'}^{\dagger}\Big]\,e^{-i(p\cdot x+\pi\cdot \theta+p\cdot x'+\pi\cdot \theta')} \nonumber \\
&& \,+\,\Big[\hat{a}_{p,\pi}^{\dagger},\hat{a}_{p',\pi'}\Big]\,e^{i(p\cdot x+\pi\cdot \theta-p\cdot x'-\pi\cdot \theta')}
\,-\,\Big[\hat{a}_{p,\pi}^{\dagger},\hat{a}_{p',\pi'}^{\dagger}\Big]\,e^{i(p\cdot x+\pi\cdot \theta+p\cdot x'+\pi\cdot \theta')} \biggrb\} \; ,
\eaa

\noindent  and using Eqs. (\ref{SS1}) and (\ref{SS2}) we have that

\baa
\label{UU}
&&\[\, \hat{\phi}(x,\theta,t), \hat{\pi} (x',\theta^{'},t)\,\]=
\,i\,\int \,d^{\,9} P\,d^{\,9} P' \,N_{p,\pi} \, N_{p'}\,(i\omega_{p'})\, \times
\nonumber \\
&&\times \, \delta^3 ({\bf p}-{\bf p}^{\prime})\,\delta^6 (\pi - \pi') \, \Big[ e^{-i[p\cdot(x-x')+\pi\cdot(\theta-\theta')]}\,+\,e^{i[p\cdot(x-x')+\pi\cdot(\theta-\theta')]} \Big] \; .
\eaa

\noindent If we choose the normalization constant

\bbe
\label{V}
N_{p,\pi}\,=\,\frac{1}{\sqrt{2\omega_{p,\pi} (2\pi)^9}} \; ,
\eee

\noindent we will obtain the result in (\ref{UU}) as

\bbe
\label{X}
\[\, \hat{\phi}(x,\theta,t), \hat{\pi} (x',\theta^{'},t)\,\]\,=\,\frac i2\,\int \frac{d^{\,9} P}{(2\pi)^9}\Big[ e^{-i[p\cdot(x-x')+\pi\cdot(\theta-\theta')]}\,+\,
e^{i[p\cdot(x-x')+\pi\cdot(\theta-\theta')]} \Big] \; ,
\eee

\noindent and finally we will have that

\baa
\label{ZZ}
\[\, \hat{\phi}(x,\theta,t), \hat{\pi} (x',\theta^{'},t)\,\]\,&=&\,\frac i2\,\int \frac{d^{\,3} {\bf p}}{(2\pi)^3}\,\frac{d^{\,6} \pi}{(2\pi)^6}
\Big[ e^{ip\cdot(x-x')}\cdot e^{i\pi\cdot(\theta-\theta')}\,+\,e^{-ip\cdot(x-x')}\cdot e^{-i\pi\cdot(\theta-\theta')} \Big] \nonumber \\
\mbox{} \nonumber \\
&=&i\,\delta^3\,({\bf x}-{\bf x}')\,\delta^6\,(\theta-\theta') \; ,
\eaa

\noindent which is the path opposite direction followed in \cite{amorim} where the delta functions are assumed to have the form obtained in (\ref{ZZ}).  Notice that what we have done here was to demonstrate the commutation relation in Eqs. (\ref{CC2}) and (\ref{CC1}) using the fields constructed with DFR phase-space definitions.

We can see that the result in (\ref{ZZ}) corroborates the construction of the operator in Eq (\ref{B}) with a convenient choice for the normalization and obeying the commutation operators.   We believe that this formalism completes both ones depicted in \cite{saxell} and \cite{ccz} since in the first one the existence of a NC six-dimensional phase-space is missing since we have shown that the existence of a momentum is connected to Lorentz invariance.  Concerning \cite{amorim}, the path here was different since we have demonstrated here that the field operator in a NC spacetime can be written in terms of lane waves in NC spacetime that can be written as

\baa
\label{Z1}
u_{p,\pi} (x,\theta t)&=&N_{p,\pi} \, e^{-i( p\cdot x + \pi \cdot \theta )}=\frac{e^{-i( p\cdot x + \pi \cdot \theta )}}{\sqrt{2\omega_{p,\pi} (2\pi)^9}} \; ,
\eaa

\noindent when we substitute Eq. (\ref{Z1}) in the Fourier expansion in Eq. (\ref{QQ}).  And the same can accomplished for $\pi_{\mu\nu}$.

The Lagrangian density of a real spin-0 field $\phi(x)=\phi(\hat{x},\hat{\theta},t)$ with mass $m$ can be written as \cite{amorim}
\bbe
\label{DD}
{\cal L}\,=\, \frac12 \, \partial^{\mu} \phi \, \partial_{\mu} \phi 
\,+\, \frac{\lambda^2}{4} \, \partial^{\mu\nu} \phi \, \partial_{\mu\nu} \phi \,-\, \frac 12 \, m^2 \phi^2 \; ,
\eee
\noindent where $\partial_{\mu\nu}:=\frac{\partial}{\partial\theta^{\mu\nu}}$, and $\lambda$ is a parameter with dimension of length, as the Planck length.
It is introduced here due to dimensional needs. From (\ref{DD}), the Klein-Gordon equation is
\bbe
\label{EE}
\( \Box \,+\,\lambda^2 \Box_{\theta}\,+\,m^2 \) \phi \,=\,0 \; ,
\eee
\noindent where $\Box =\partial^{\mu}\partial_{\mu}$ and $\Box_{\theta} =\frac 12 \partial^{\mu\nu}\partial_{\mu\nu}$ is the four- and six-dimensional Laplace operators, respectively.  The canonical conjugate momentum is
\bbe
\label{FF}
\pi(x)\,=\,\frac{\partial {\cal L}}{\partial \dot{\phi} (x)}\,=\,\dot{\phi}(x) \; ,
\eee
\noindent in which $\pi(x)=\pi(x,\theta,t)$. This result leads us to the Hamiltonian density
\baa
\label{GG}
{\cal H}&=& \pi(x) \dot{\phi}(x)\,-\,{\cal L} \nonumber \\
&=& \frac 12 \[ \pi^{2}(x) \,+\,(\nabla\,\phi(x))^2\,+\,(\lambda\,\nabla_{\theta} \phi(x))^2\,-\,m^2 \phi^{2}(x) \]
\eaa
\noindent where $\nabla_{\theta}=\frac 12 \partial^{ij}$, and the quantized Hamiltonian operator is given by
\bbe
\label{HH}
\hat{H}\,=\,\int\,d^3 {\bf x} \, d^6 \theta\,\frac 12 \Big[ \hat{\pi}^{2}(x,\theta,t)\,+\,(\nabla\hat{\phi} (x,\theta,t))^2\,+\,( \lambda \nabla_{\theta} \hat{\phi} (x,\theta,t))^2\,-\,m^2 \hat{\phi}^{2} (x,\theta,t) \Big] \; .
\eee
\noindent We can use the commutation relations in Eqs, (\ref{CC1}) and (\ref{CC2}) to calculate the Hamilton's equations of motion as
\bbe
\label{II1}
\dot{\hat{\phi}} (x,\theta,t)\,=\,-i\[\, \hat{\phi} (x,\theta,t), \hat{H} \,\]\,=\,\hat{\pi} (x,\theta,t) \; ,
\eee
\noindent and
\bbe
\label{II2}
\dot{\hat{\pi}}\,=\,-i\[ \, \hat{\pi} (x,\theta,t), \hat{H}\, \] \,=\, \Big( \nabla^2\,+\,\lambda^2 \nabla^2_{\theta}\,-\,m^2 \Big)\, \hat{\phi} (x,\theta,t) \; ,
\eee

\noindent where we have integrated by parts where it was needed.  Notice that, using  Eqs. (\ref{II1}) and (\ref{II2}) we can construct the NC Klein-Gordon equation

\bbe
\label{JJ}
\ddot{\hat{\phi}}(x,\theta,t)\,=\,\Big(\nabla^2\,+\,\lambda^2 \nabla^2_{\theta}\,-\,m^2 \Big)\,\hat{\phi}(x,\theta,t) \; ,
\eee

\noindent which shows clearly a different path from \cite{saxell} since the author did not consider the existence of a canonical momentum.


\section{Conclusions}

The investigation of the physical theories that happen in NC space-time has brought great interest through the last years and one of the reasons for that interest is the hope to understand gravity at the Planck scale.   The existence of a parameter that allows the comprehension of NC theories as laboratories to study the physics of the very early Universe motivates theoretical physicists to pursue this NC knowledge.  In other words, we hope to find an algebraic unified model \cite{filk} or a arguably understanding of a quantum space as the beginning of a quantum gravity theory which avoids current problems and to be free of singularities, for instance.   The seminal objective would be that the deformation of space-time would act as a regularization scheme which would keeps the algebraic properties of the theory.

The extended DFR formulation of NC theories was developed recently and its main mathematical characteristic is to promote the NC parameter, $\theta^{\mu\nu}$, to the status of space-time coordinates.  This procedure recovers the Lorentz invariance of the theory and at the same time it requires
the construction of a conjugated momentum associated with $\theta^{\mu\nu}$, together with its respective algebra.  In this work, based on the results obtained in two different $\theta$-variable phase-spaces, we have shown, in the DFR space-time, that this momentum $\pi_{\mu\nu}$ (which completes the set of phase-space symplectic variables as being $(x^{\mu},p_{\mu}, \theta^{\mu\nu}, \pi_{\mu\nu})$) is directly connected to Lorentz invariance and cannot be considered irrelevant in any ordinary DFR analysis since it is essential to calculate the QFT commutation relations for DFR formalism.

Through two examples, the DFR harmonic oscillator and the NC relativistic particle developed in \cite{alexei}, we have shown that in both NC formalisms (a DFR algebra and a non-DFR algebra, respectively) we have a kind of duality $\theta_{constant} \longrightarrow  \theta_{variable}$ which can be also represented by $\pi=0 \longrightarrow \pi \not= 0$ or (non-Lorentz invariance) $\longrightarrow$ (Lorentz invariance) maps.   The conclusion is that there is no difference between DFR and DFR$^*$, and consequently the DFR formalism has the conjugated pairs $(x,p)$ and $(\theta,\, \pi)$.   We believe that this result complements the DFR literature.   In this way we have constructed also the scalar field QFT and we have calculated the operatorial commutation relations with the $(x,p,\,\theta,\,\pi)$ phase-space.

The NC relativistic particle shows, besides the $\theta_{constant} \longrightarrow \theta_{variable}$ duality, another interesting result.   Since the equations of motion have shown that for $\theta=\mbox{constant}$ we have the multiplier $\lambda_{\theta} = 0$ and this value zeroes the NC acceleration, the velocity is not constant since it has a parameter that is time dependent.
In \cite{alexei} the author has obtained this last result also, but since he does not have the value of the acceleration, it was not possible to see how interesting this result is. Besides, we have calculated here that $\dot{e}\not= 0$, which confirms that, following the equations of motion, the velocity $\dot{x}$ is not constant. It was important to compute $\dot{e}$ because although it is defined as $e=e(\tau)$ its calculation could result as zero, which would show a paradox, but it did not happen.

As a perspective we can analyze other $\theta_{variable}$ algebras different from DFR (of course) to verify if the behavior is the same. Another possible research is to construct the fermion DFR QFT. It is an ongoing research and it will published elsewhere.


\begin{acknowledgments}
\ni E.M.C.A. thanks CNPq (Conselho Nacional de Desenvolvimento Cient\' ifico e Tecnol\'ogico), a Brazilian research support federal agency, for partial financial support.
\end{acknowledgments}

\appendix

\section{The Moyal-Weyl product}

To investigate field theories defined on spaces with noncommutative coordinates corresponding to deformations of flat spaces as e.g. the Euclidean plane or Minkowski space $\mathbb{M}^d$ one must replace the (commuting) coordinates of flat space by Hermitian operators $x^\mu$ (with $\mu$ = 0, 1, $\cdots$ , ($d$-1)) \cite{filk}. We consider a canonical structure defined by the following algebra

\baa
\label{filk}
\left[ \hat{x}^\mu,\hat{x}^\nu \right] =i\theta^{\mu\nu}
\; , \;
\left[ \theta^{\mu\nu},\hat{x}^\rho\right] =0 \,\,.
\eaa

\ni The simplest case is when the $\theta^{\mu\nu}$ matrix is constant, which means that we have only the first equation of (\ref{filk}). Furthermore, it is real and antisymmetric. In natural units, where $\hbar = c = 1$, it can be seen easily from (\ref{filk}) that it has squared mass dimension.

In order to construct the perturbative field theory formulation, it is more
convenient to use fields $\Phi(x)$ (which are functions of ordinary commuting coordinates) instead of operator valued objects like $\hat{\Phi}(\hat{x})$. To be able to pass to such fields, in respecting the properties (\ref{filk}), one must redefine the multiplication law of functional (field) space. One therefore defines the linear map $\hat{f}(\hat{x})\longmapsto S[\hat{f}](x) $, called the symbol of the operator $\hat{f}$, and can then represent the original operator multiplication in terms of so-called star products of symbols as
\bbe
\hat{f}\hat{g}=S^{-1}\left[ S[\hat{f}]\star S[\hat{g}]\right] \,\,,
\eee
\ni see for example references \cite{douglas,szabo}. By using the Weyl-ordered symbol (which corresponds to the Weyl-ordering prescription of the operators) one can arrive at the following definitions, with $S[\hat{f}](x)=\Phi(x)$, we have
\baa
\label{weyl}
\hat{\Phi}(\hat{x})&\longleftrightarrow&\Phi(x), \no \\
\hat{\Phi}(\hat{x})\hspace{-0.3cm}&=& \hspace{-0.3cm} \int \frac{d^dk}{(2\pi)^d}\enspace \tilde{\Phi}(k) \, \, e^{ik\cdot\hat{x}}, \no \\
\tilde{\Phi}(k)\hspace{-0.3cm}&=& \hspace{-0.3cm} \int d^dx\enspace \Phi(x) \, \, e^{-ik \cdot x}\,\,,
\eaa
\ni where $k$ is real variable, and $\hat{x}$ is the position operator. For any two arbitrary scalar fields $\hat{\Phi}_1$ and $\hat{\Phi}_2$ one therefore can write that \footnote{One has to use the Baker-Campbell-Hausdorff formula, as well as relation (\ref{filk})}
\baa
\label{star}
\hat{\Phi}_1(\hat{x}) \, \hat{\Phi}_2(\hat{x})&=&\int\frac{d^dk_1}{(2\pi)^d}\int\frac{d^dk_2}{(2\pi)^d} \, \, \tilde{\Phi}_1(k_1) \, \tilde{\Phi}_2(k_2)\enspace e^{ik_1\cdot\hat{x}} \enspace e^{ik_2\cdot\hat{x}} \no \\
&=& \int\frac{d^dk_1}{(2\pi)^d}\int\frac{d^dk_2}{(2\pi)^d} \, \, \tilde{\Phi}_1(k_1) \, \tilde{\Phi}_2(k_2)\enspace e^{i(k_1+k_2)\cdot\hat{x}-\frac{1}{2}[\hat{x}^{\mu},\hat{x}^{\nu}]k_{1\mu}k_{2\nu}} \; .
\eaa

\ni Hence one has the following Weyl-Moyal correspondence \cite{bgps}

\bbe
\hat{\Phi}_1(\hat{x}) \, \hat{\Phi}_2(\hat{x}) \longleftrightarrow \Phi_1(x) \star \Phi_2(x) \; ,
\eee

\ni where, in using relation (\ref{filk}) to replace the commutator in the exponent of (\ref{star}), the generalized Moyal-Weyl star product is given by

\bbe
\label{gmw}
\Phi_1(x) \star \Phi_2(x)\, = \, \mbox{exp}\Big(\frac i2 \, \theta^{\mu\nu} \partial_{\mu x} \partial_{\nu y} \Big) \, \Phi_1(x) \, \Phi_2(y)\Big|_{x=y} \; .
\eee

\ni This means that we can work in the same way as in a usual commutative space for which the
multiplication operation is modified by the star product (\ref{gmw}). For the ordinary commuting coordinates this implies\footnote{The Weyl bracket is defined as $[A,B]_{\star}=A\star B-B\star A$}
\baa
\left[ x^{\mu},x^{\nu}\right] _{\star}= i\theta^{\mu\nu}
\; \; , \; \;
\left[ \theta^{\mu\nu}, x^{\rho}\right] _{\star}=0\,\,.
\eaa

\ni At this point one also has to mention that the commutation relations
(\ref{filk}) between the coordinates explicitly break Lorentz invariance because of the fact that we assumed $\theta$ is a constant matrix \cite{chklo}.

Some other possibilities for a  non-constant $\theta$ are, for example, $\theta^{\mu\nu}=C^{\mu\nu}_{\;\;\;\;\rho} \, x^{\rho}$ (Lie algebra) or $\theta^{\mu\nu}=R^{\mu\nu}_{\;\;\;\;\rho\sigma} \, x^{\rho} \, x^{\sigma}$ (quantum space structure) - see for instance reference \cite{douglas,szabo} for a detailed discussion about these two approaches.

Another solution of this problem leads us to the NC formulation of the spacetime used here which was formulated by Doplicher, Fredenhagen and Roberts (DFR) \cite{dfr}, which is based in general relativity and quantum mechanics arguments.  This formalism recovers Lorentz invariance through the promotion of $\theta^{\mu\nu}$ to be a standard coordinate of this extra dimensional system.  Of course, being the coordinate, the algebra turns out to be, together with Eq. (\ref{filk})

\bbe
\label{Capp}
\[ \hat{x}^{\mu} , \hat{p}^{\nu} \]\,=\, i \eta^{\mu\nu} \; \; ,
\; \; \[\hat{x}^{\mu} , \hat{\theta}^{\mu\nu} \]\,=\,\[\hat{p}^{\mu} , \hat{p}^{\nu}\]\,=\,\[ \hat{\theta}^{\mu\nu} , \hat{\theta}^{\rho\lambda}\]\,=\,0 \; ,
\eee

\noindent which completes the basic DFR algebra.

\bigskip
\bigskip



\end{document}